\documentclass[pre,aps,twocolumn,showpacs]{revtex4}
\usepackage{amsmath,amssymb,graphicx}
\usepackage{epsfig}
\usepackage{textcomp}
\usepackage{color}

\begin{document}
\title{Immigration-extinction dynamics of stochastic populations}

\author{Baruch Meerson$^{1}$ and Otso Ovaskainen$^2$}

\affiliation{$^{1}$Racah Institute of Physics, Hebrew University
of Jerusalem, Jerusalem 91904, Israel}
\affiliation{$^{2}$Department of Biosciences, University of Helsinki, FI-00014 Helsinki, Finland}


\begin{abstract}
How high should be the rate of immigration into a stochastic population in order to significantly reduce the probability of observing the population extinct? Is there any relation between the population size distributions with and without immigration? Under what conditions can one justify the simple \emph{patch occupancy models} which ignore the population distribution and its dynamics in a patch, and treat a patch simply as either occupied or empty? We address these questions by exactly solving a simple stochastic model obtained by adding a steady immigration to a variant of the Verhulst model: a prototypical model of an isolated stochastic population.

\end{abstract}

\maketitle

\section{Introduction}

Any isolated population ultimately goes extinct with probability one.
This happens,  even in the absence of detrimental environmental variations,
because of the shot noise of elementary processes of reproduction and death. The noise (also called demographic
stochasticity)
ultimately brings the population to an absorbing state at zero population size \cite{Bartlett,Nisbet,OM}.  In sufficiently large, and therefore long-lived populations, extinction is caused by
a rare chain of events when population losses dominate over gains. In this case the probability distribution function (PDF) of population size exhibits a long-lived metastable, or quasi-stationary state, and
the probability is slowly ``leaking" into the absorbing state, see Ref. \cite{OM}
and references therein.  A mathematically similar setting appears in many other contexts in physics, chemistry, biology and other fields.  One important example is extinction of a disease from a population following an endemic, under condition that no new infectives arrive \cite{Bartlett}.

It has long been recognized that immigrants from surrounding populations can prevent local extinction of a small population by either recolonization of empty regions or the ``rescue effect" \cite{Brown,Hanski}.
Similarly, introduction of new infected individuals into a population, which has recovered from an infection, can reignite the epidemics.  Mathematically, by introducing a steady immigration flux, with no matter how small a rate, one eliminates the absorbing state at zero population size and therefore prevents extinction. At too a low immigration rate, however, this prevention is only nominal. Indeed, if the immigration rate is much lower than the characteristic extinction rate of  the isolated population, the observed population size will most likely be zero for any chosen moment of time.  How significant should the immigration rate be so that  the probability of observing a local population extinct is considerably reduced? This question was already posed by MacArthur and Wilson \cite{MacArthur,MacArthurbook} who discussed how the equilibrium number of species (the sum of species-specific occupancies), found on an island, depends on the balance of colonization and extinction rates. More recently, Matis and Kiffe \cite{Matis} considered a nonlinear stochastic population model with immigration and suggested a cumulant truncation procedure to approximate the PDF of the population size in this model. We will deal with a similar model here, but solve it exactly. Furthermore, we use the exact solution to address two additional questions that have not been addressed before. The first question is the following: Is there any connection between (i) the truly stationary PDF of the population size in the case with immigration and (ii) the quasi-stationary PDF of the population size without immigration? The second question is: under what conditions can one justify the simple \emph{patch occupancy models} \cite{SPOM}  which ignore the population distribution and its dynamics in a patch, and treat a patch simply as either occupied or empty?

The model we will be dealing with is obtained by adding a steady immigration process to a variant of the Verhulst model: a prototypical model of isolated stochastic population.  In Sec. 2 we will first outline the Verhulst model without immigration, and then add immigration process. In Sec. III we derive the exact solution for the PDF of the population size. Sec. IV presents the results of asymptotic analysis of the exact solution in the regime of low immigration rate and large mean population size.
In Sec. V we discuss the connection between our Verhulst model with immigration and a simple two-state stochastic patch occupancy model. In Sec. VI we
extend some of our finding to more complicated situations. The main results are briefly summarized in Sec. VII.

\section{Model}

\subsection{Without immigration}
We assume a well-mixed single population where individuals undergo reproduction and death. In the absence of immigration, the model coincides with a variant of Verhulst model \cite{Nasell,Doering,AM}. The reproduction and mortality
rates are given by
\begin{equation}\label{verhulst}
\lambda_n= B\,n\;\;\;\mbox{and}\;\;\;\mu_n=n+\frac{Bn^2}{N}\,,
\end{equation}
respectively, and time and the rates are rescaled by the linear in $n$ term in the mortality rate.  We will assume throughout most of this paper that the reproduction rate of the population is greater than one: $B>1$. In this case the deterministic rate equation,
\begin{equation}\label{rateeq0}
\frac{d \bar{n}}{dt}= (B -1)\,\bar{n}(t)-\frac{B}{N}\, \bar{n}^2(t) \,.
\end{equation}
has an attracting fixed point, $\bar{n}=N(1-1/B)$. This fixed point describes an established population which, according to the deterministic theory, persists forever. In the stochastic model the shot noise drives the population to extinction. Correspondingly, the stationary PDF of the population size is equal to the Kronecker's delta $\delta_{n,0}$, so the probability  to find the system empty at  $t\to \infty$ is equal to 1. For $N \gg 1$, the mean time to extinction is exponentially long in $N$:
\begin{equation}\label{MTE}
    \tau \simeq \sqrt{\frac{2\pi}{N}}\,\frac{\sqrt{B}}{(B-1)^2}\, e^{N \left(1 -\frac{1}{B}-\frac{\ln B}{B}\right)}\,,
\end{equation}
see Refs. \cite{Nasell,Doering,AM}. The quasi-stationary PDF of the population size can be calculated from the general formulas of Ref. \cite{AM}. In the limit of $N\gg 1$ and $n\gg 1$ the (normalized to unity) quasi-stationary PDF is
\begin{equation}\label{qsd}
    \pi_n\simeq\frac{(B-1) N e^{-N \left[1-\frac{1}{B}-\frac{n}{N}+\left(\frac{n}{N}+\frac{1}{B}\right) \ln \left(\frac{n}{N}+\frac{1}{B}\right)\right]}}{n \sqrt{2\pi B (N+Bn)}}\,.
\end{equation}

\subsection{With immigration}

Now we add to the model steady immigration: arrival of new individuals with rate $r>0$ \cite{rescaled}. The stationary PDF of the population size is now non-trivial. In particular, no matter how small $r$ is, the probability, $P_0$,  of observing the system empty at $t \to \infty$ is less than one. How does $P_0$ depend on the immigration rate $r$? Furthermore, is there any relation between the stationary PDF $P_n$ observed at $r>0$ and the quasi-stationary PDF $\pi_n$ observed for $r=0$?

The deterministic rate equation becomes
\begin{equation}\label{rateeq}
\frac{d \bar{n}}{dt}= r+ (B -1)\,\bar{n}(t)-\frac{B}{N}\, \bar{n}^2(t) \,.
\end{equation}
This equation has an attracting fixed point $n_s$ such that
\begin{equation}\label{fp}
    \frac{\bar{n}_s}{N}=\frac{B-1+\sqrt{(B-1)^2+4 Br/N}}{2 B}\,.
\end{equation}
The expected value of the population size in the stochastic model is expected to be peaked, in the leading order in $N\gg 1$, at $\bar{n}_s$.

A complete probabilistic description of the Verhulst model with immigration is provided by the continuous-time master equation
\begin{eqnarray}
&&\frac{dP_n}{d t} = r P_{n-1}-r P_{n}
+B (n-1)P_{n-1} - Bn P_n \nonumber\\
&+& \left[n+1+\frac{B (n+1)^2}{N}\right] P_{n+1}
-\left(n+\frac{Bn^2}{N}\right)P_n  ,\label{master}
\end{eqnarray}
where $P_n(t)$ is the probability of observing the population size $n$ ($n=0,1,\dots$) at
time $t$.

\section{Exact solution for the steady-state PDF}

Let us introduce the probability generating function \cite{vK}
\begin{equation}
G(p,t)=\sum_{n=0}^{\infty} p^n P_n(t)\,,
\end{equation}
where $p$ is an auxiliary variable. $G(p,t)$ obeys the
normalization condition
\begin{equation}\label{norm}
    G(1,t)=1
\end{equation}
which follows from the conservation of the total probability. The probability generating function encodes all the
probabilities $P_n(t)$, as those are given by the coefficients
of the Taylor expansion of $G(p,t)$ around $p=0$:
\begin{equation}\label{prob}
P_n(t)=\left.\frac{1}{n!}\frac{\partial ^n G}{\partial
p^n}\right|_{p=0}\,.
\end{equation}
In their turn, the moments of the PDF can be expressed through the $p$-derivatives
of the generating function at $p=1$, \textit{e.g.} $\langle
n\rangle(t) \equiv \sum_n n P_n(t) = \left.\partial_p
G(p,t)\right|_{p=1}$.

By multiplying Eq.~(\ref{master}) by $p^n$ and summing over all
$n$ one obtains an evolution equation for
$G(p,t)$:
\begin{eqnarray}\label{Gdot1}
\frac{\partial G}{\partial t} &=&
\frac{B}{N}\left(1-p\right)p\,\frac{\partial^2 G}{\partial
p^2}\nonumber \\
&+&(p-1)\left(Bp-1-\frac{B}{N}\right)\frac{\partial
G}{\partial p}+r(p-1) G \,.
\end{eqnarray}
At sufficiently long times, a stationary PDF $G_0(p)$ sets in. It is described by the second-order ODE
\begin{equation}\label{ode}
\frac{B}{N} p \,G_0^{\prime\prime}(p)-\left(B p-1-\frac{B}{N}\right) G_0^{\prime}(p)-r G_0(p)=0\,,
\end{equation}
where primes denote the derivatives with respect to the argument. Equation (\ref{ode}) is the Laguerre differential equation. One of its two independent solutions blows up at $p=0$ and, in view of Eq.~(\ref{prob}),  must be discarded. The other solution is well behaved. Choosing the arbitrary constant so as to satisfy the normalization condition (\ref{norm}), we obtain
\begin{equation}\label{Gexact1}
    G_0(p)=\frac{\, _1F_1\left(\frac{r}{B};\frac{B+N}{B};N p\right)}{\,
   _1F_1\left(\frac{r}{B};\frac{B+N}{B};N\right)}\,,
\end{equation}
where $\, _1F_1 (a;b;z)$ is the  Kummer confluent hypergeometric function \cite{Abramowitz}. If there is no immigration, $r=0$, Eq.~(\ref{Gexact1}) yields $G_0(p)=1$ which corresponds to an empty system, $P_n=\delta_{n,0}$, as expected.

\begin{figure}[ht]
\includegraphics[width=2.5in,clip=]{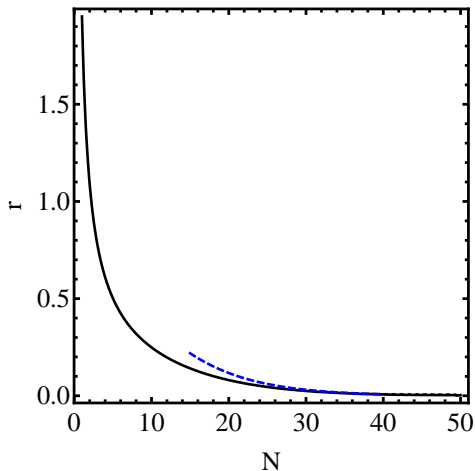}
\caption{(Color online) The minimum value of the immigration rate $r$ versus $N$, needed for bringing the probability of observing the population extinct below $0.5$ as given by Eq. (\ref{P0}) (solid line) and by the asymptotic (\ref{finalP0}) (dashed line). Parameter $B=2$.} \label{pd}
\end{figure}

Now we can use Eq.~(\ref{prob}) to find the stationary PDF $P_n$. It is convenient to perform the symbolic differentiation with ``Mathematica" \cite{mathematica}, obtaining
\begin{equation}\label{stpdf}
P_n=\frac{N^n \,\Gamma \left(1+\frac{N}{B}\right)
\Gamma\left(n+\frac{r}{B}\right)}{n! \,\Gamma
\left(\frac{r}{B}\right) \Gamma
   \left(n+\frac{N}{B}+1\right)\,_1 F_1 \left(\frac{r}{B};1+\frac{N}{B};N\right)}\,,
\end{equation}
where $\Gamma(z)$ is the gamma function \cite{Abramowitz}. In particular,  the probability to find the system empty is
\begin{equation}\label{P0}
    P_0=\frac{1}{\, _1F_1\left(\frac{r}{B};1+\frac{N}{B};N\right)}\,.
\end{equation}
Equations (\ref{stpdf}) and (\ref{P0}) are exact and give the complete stationary PDF for any positive values
of $N$, $B$ and $r$. Equation (\ref{P0}) enables us  to determine the minimum value of the immigration rate $r$ needed, at given $N$ and $B$, for bringing the probability of observing the population extinct  below, say,  $0.5$. An example
is given in Fig. \ref{pd}.

\section{Low immigration rate}
\begin{figure}[ht]
\includegraphics[width=3in,clip=]{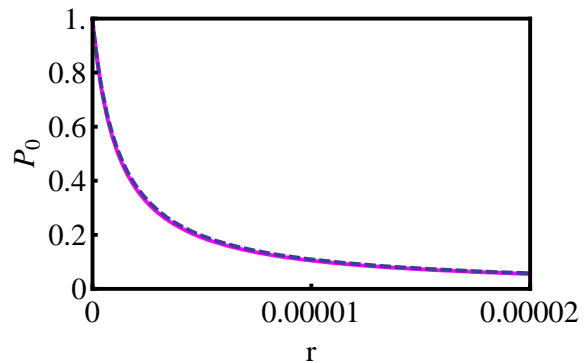}
\caption{(Color onlie) Solid line: the exact stationary probability to find the system empty versus $r$, see Eq.~(\ref{P0}). Dashed line: approximation~(\ref{finalP0}). The parameters are $N=100$ and $B=2$, so that $\tau\simeq 1.63 \times 10^6$. }  \label{P0fig}
\end{figure}

Most interesting for our purposes is the regime of low immigration rate and a large population size. Here one can simplify Eqs. (\ref{stpdf}) and (\ref{P0}) by employing the strong inequalities  $N\gg 1$ and $r\ll B$  and using asymptotic expansions of the special functions, see Appendix. For the stationary probability $P_0$ to find the system empty we obtain the following approximation:
\begin{equation}\label{finalP0}
    P_0 \simeq \frac{1}{1+\left(1-\frac{1}{B}\right) r\tau}\,,
\end{equation}
where $\tau$, given by Eq.~(\ref{MTE}), is the mean time to extinction of the same population but without immigration (that is, at $r=0$). As expected, $P_0 \to 1$ as $r \to 0$. When $(1-1/B) r\tau$ is much larger than $1$, $P_0$ becomes exponentially small (because of the presence of $\tau$ in the denominator) and goes down
as $1/r$ as $r$ increases. Figure \ref{pd} illustrates that the approximation (\ref{finalP0}) becomes accurate at sufficiently large $N$ and small $r$. In its turn, Fig.~\ref{P0fig} compares the exact stationary probability $P_0$ versus $r$, see Eq.~(\ref{P0}), with the approximation (\ref{finalP0}), for $N=100$ and $B=2$, and good agreement is observed.

For $P_{n\gg1}$ the asymptotic expansion (see Appendix) yields
\begin{equation}\label{Pnappr}
   P_{n\gg1}\simeq \frac{r
\,e^{N\left[\frac{n}{N}-\frac{\ln B}{B}-\left(\frac{n}{N}+\frac{1}{B}\right) \ln\left(\frac{n}{N}+\frac{1}{B}\right)\right]}}{B n \left(1+\frac{Bn}{N}\right)^{1/2}\left[1+\left(1-\frac{1}{B}\right) r\tau\right]}\,.
\end{equation}
This result holds for $N\gg 1$, $n\gg 1$ and $r\ll B/\ln N$. As one can check by inspection, Eq.~(\ref{Pnappr}) can
be rewritten as
\begin{equation}\label{Pnappra}
P_{n\gg 1}\simeq \frac{\left(1-\frac{1}{B}\right) r\tau}{1+\left(1-\frac{1}{B}\right) r\tau}\,\pi_n,
\end{equation}
where $\pi_n$ is the quasi-stationary PDF of the population without immigration, see Eq. (\ref{qsd}),
and $\tau$ is the mean time to extinction of that population, see Eq. (\ref{MTE}).
When $(1-1/B)\, r \tau\ll 1$ the PDF,
\begin{equation}\label{Pnapprb}
 P_{n\gg 1}\simeq\left(1-\frac{1}{B}\right) r\tau\,\pi_n,
\end{equation}
is exponentially small. Here almost all of the probability is concentrated in the empty state, $P_0$. As $r \to 0$, $P_{n\gg 1} \to 0$ as expected. For $(1-1/B)\, r\tau \gg 1$ (but still $r \ll B/\ln N$) we obtain
\begin{equation}\label{Pnappr1}
    P_{n\gg 1}\simeq\pi_n\,.
\end{equation}
As one can see from this equation, in a wide range of (sufficiently small, but not too small) immigration rates, the probability of observing a significant population size is independent of the immigration rate and coincides with the quasi-stationary PDF (\ref{qsd}) of the same population but without immigration.

PDF (\ref{Pnappr}) has its maximum at
$$
n=n_*=N\left(1-\frac{1}{B}\right)-\frac{3B-1}{2 (B-1)}+{\cal O}(1/N)\,,
$$
that is very close to the $N(1-1/B)$, the attracting fixed point~(\ref{fp}) of the deterministic rate equation (\ref{rateeq}) for a zero immigration rate $r=0$. Furthermore, in a vicinity of $n=n_*$ PDF (\ref{Pnappr}) can be approximated by a Gaussian:
\begin{equation}\label{gauss}
    P_{G}(n)\simeq \frac{1}{\sqrt{2\pi N}}\,\frac{\left(1-\frac{1}{B}\right) r\tau}{1+\left(1-\frac{1}{B}\right) r\tau}\,e^{-\frac{(n-n_*)^2}{2N}}\,.
\end{equation}
The Gaussian region makes a dominant contribution to the total probability ${\cal P}\equiv \sum_{n=1}^{\infty} P_n$ of observing the population non-extinct. One obtains
\begin{equation}\label{nonempty}
{\cal P}=\frac{\left(1-\frac{1}{B}\right) r\tau}{1+\left(1-\frac{1}{B}\right)},
\end{equation}
so ${\cal P}$ plus $P_0$ from Eq.~(\ref{finalP0}) yields $1$ as expected.

\begin{figure}[ht]
\includegraphics[width=3in,clip=]{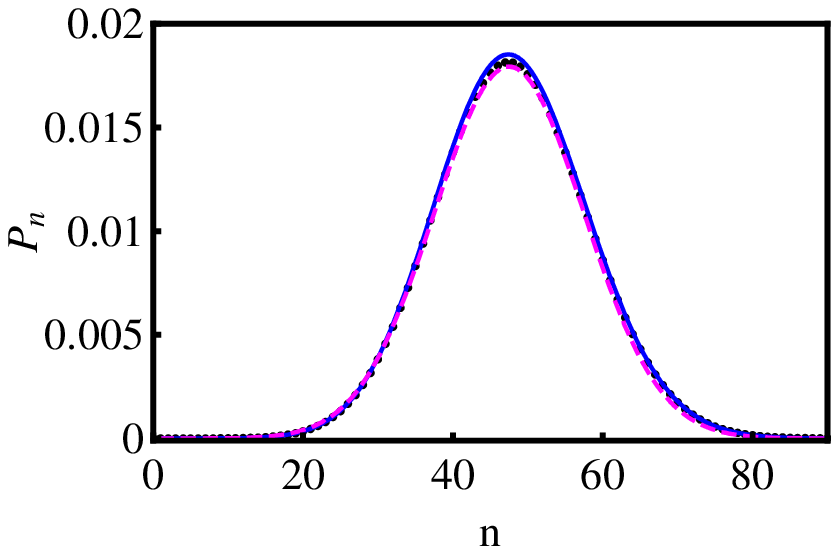}
\includegraphics[width=3.2in,clip=]{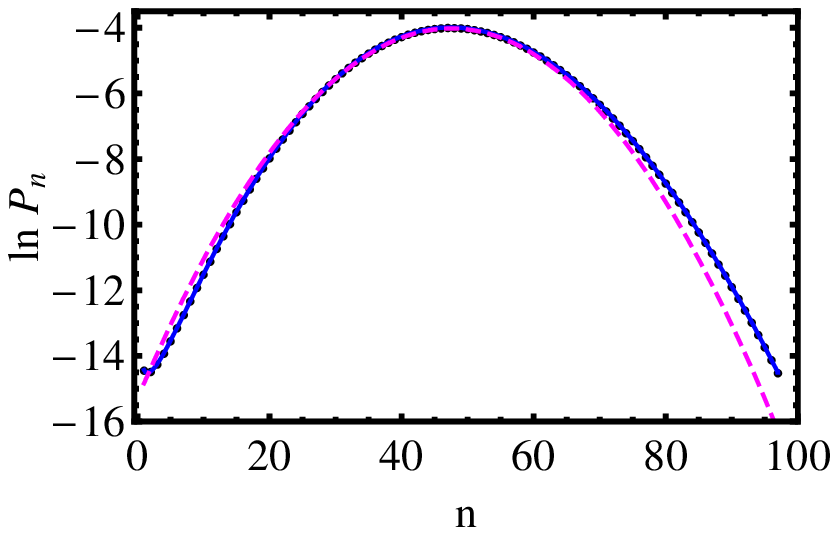}
\caption{(Color online) A comparison of the exact and approximate stationary PDFs  for $N=100, B=2$ and $r=10^{-6}$. Circles: exact results. Solid curve: Eq.~(\ref{Pnappr}). Dashed curve: the Gaussian asymptotic (\ref{gauss}).
The lower panel shows the natural logarithm of the corresponding distributions.}  \label{Pnfig}
\end{figure}

Figure~\ref{Pnfig} shows a comparison of the approximate stationary PDF (\ref{Pnappr}) with the exact result (\ref{stpdf}) for $N=100, B=2$ and $r=10^{-6}$. As one can see, the agreement is excellent. For these parameters the stationary probability to find the population extinct is $P_0=0.5393\dots$, whereas the approximation Eq.~(\ref{finalP0}) predicts
$P_0\simeq 0.55$. Also shown in Fig. ~\ref{Pnfig} is the Gaussian asymptotic (\ref{gauss}). It agrees well with the exact result in the ``body"  of the PDF. It is not as faithful, however, in the distribution tails, as evidenced by the lower panel of  Fig. \ref{Pnfig} which depicts the corresponding distributions in logarithmic scale.

\section{Patch occupancy model}
In metapopulation theory, much of work has been based on stochastic patch occupancy models \cite{SPOM} which ignore the local population distribution and its dynamics in a patch and treat a patch simply as either occupied or empty. The simplest stochastic patch occupancy model appears in the context of mainland-island model \cite{MacArthur}, where the mainland effectively serves as a stationary reservoir of immigrants. As we will now show, in the regime where the population is long-lived and the immigration rate is low,  the results obtained with a patch occupancy model agree with the results we have obtained from the Verhulst model with immigration.

Consider a single patch (island) that can be either occupied, with probability ${\cal P}$, or  empty with probability ${\cal P}_0$, so that ${\cal P}+{\cal P}_0=1$.  Assuming that both immigration events into the patch and extinction events of the patch population are rare, and therefore independent and Poisson-distributed, we set the transition rate from the empty state to the occupied state as $\lambda=R$, and the transition rate of the reverse process as $\mu$. The dynamics of this simple two-state system is described by the equations
\begin{eqnarray}
  \frac{d{\cal P}_0}{dt} &=& -R {\cal P}_0+ \mu {\cal P}, \nonumber \\
  \frac{d{\cal P}}{dt} &=& R {\cal P}_0- \mu {\cal P}.  \label{set}
\end{eqnarray}
In the steady state one obtains
\begin{equation}\label{steadysimple}
{\cal P}_0(t\to \infty)=\frac{\mu}{R+\mu},\;\;\;{\cal P}(t\to \infty) =\frac{R}{R+\mu}.
\end{equation}
Now let us relate the rates $R$ and $\mu$ with the detailed model we dealt with in the previous sections. The transition rate $\mu$ from the occupied state to the empty state is equal to $1/\tau$: the extinction rate of the population without immigration.  One might assume that the transition rate $R$ from the empty state to the occupied state can be identified with $r$: the immigration rate of the detailed model. This assumption, however, would overestimate $R$. This is because the immigration rate $r$ of the detailed model is merely an \emph{attempt} rate of occupying the site, while the rate $R$ characterizes \emph{successful} attempts.  To find $R$ we should go back to our detailed model where, at low population sizes, we can neglect the nonlinear term in the mortality rate (\ref{verhulst}). Therefore, we set $\lambda_n=Bn$ and $\mu_n=n$ (the immigration process is ignored as $r$ is very small). Let $p_n$ be the probability of population going extinct before establishment if starting from a small number of $n$ individuals. If the population currently has $n$ individuals, the probabilities of the population size becoming $n-1$ and $n+1$ are
$$
m_{n-}=\frac{\mu_n}{\lambda_n+\mu_n}=\frac{1}{B+1}
$$
and
$$
m_{n+}=\frac{\lambda_n}{\lambda_n+\mu_n}=\frac{B}{B+1},
$$
respectively.
Hence $p_n$ satisfies the equation
\begin{equation}\label{recursive}
p_n=\frac{1}{B+1}\, p_{n-1}+\frac{B}{B+1} \,p_{n+1}.
\end{equation}
The general solution of this linear recursive equation, $p_n=C_1+C_2/B^n$, includes two arbitrary constants.
Demanding obvious boundary conditions $p_0=1$ and $p_{\infty}=0$, we arrive at the unique solution $p_n=1/B^n$ \cite{Nisbet,Goel}. In particular, upon arrival of a single individual, $n=1$, the population either goes extinct with probability $1/B$ or gets established with probability $1-1/B$. As a result, the rate of successful establishment is $R=r (1-1/B)$. With this $R$ and $\mu=1/\tau$, the stationary probabilities
${\cal P}_0$ and ${\cal P}$  from Eq.~(\ref{steadysimple}) coincide with the predictions from Eqs.~(\ref{finalP0})
and (\ref{nonempty}), respectively.

The simple patch-occupancy equations (\ref{set}) also predict the characteristic relaxation time of the system
toward the stationary state (\ref{steadysimple}):
\begin{equation}\label{reltime}
t_r = \frac{1}{R+\mu} = \frac{\tau}{1+\left(1-\frac{1}{B}\right)\,r\tau}.
\end{equation}
Note that in  this regime the relaxation time is quite long. When $r\to 0$, the relaxation time (towards the empty state) approaches $\tau$, the mean time to extinction in the system without immigration. When $(1-1/B)r \tau\gg 1$, the relaxation time is
equal to the inverse rate of successful establishment, $1/R$.  We expect that Eq.~(\ref{reltime}) also holds, for large $N$ and small $r$,  for the detailed model (\ref{master}), where it gives the relaxation time toward the stationary PDF described by Eqs. (\ref{finalP0}) and (\ref{Pnappr}) [or (\ref{Pnappra})].

\section{Some extensions}

The connection to the patch occupancy model provides a simple, albeit non-rigorous, tool for extending some of our results to more complex situations that those accounted for by the simple Verhulst model. (We should remember, however, to demand a sufficiently low immigration rate, and that the population without immigration is long-lived.) One immediate extension is to allow the immigrants to arrive in groups. Let $q_n$ be the probability that the arriving group includes $n$ individuals, $n=1, 2, 3, \dots$.  Then the probability that an arriving group leads to successful establishment is $R_g=\sum_n q_n (1-p_n)$, where $p_n =1/B^n$, see the paragraph after Eq.~(\ref{recursive}). For example, let the group size be Poisson distributed with mean $\nu$,  $q_n=e^{-\nu} \nu^n/n!$, and let the arrival rate of groups be $r/\nu$ \cite{rescaled} so that the per-capita arrival rate is $r$ as before. It is easy to see that the rate of successful establishment in this case is
\begin{equation}\label{poissonz}
R_g=\frac{r \left[1- e^{-(1-1/B)\,\nu} \right]}{\nu},
\end{equation}
with corresponding results for the steady state values
\begin{equation}\label{steadygroups}
{\cal P}_0(t\to \infty)=\frac{1}{1+R_g \tau},\;\;\;{\cal P}(t\to \infty) =\frac{R_g \tau}{1+R_g \tau}.
\end{equation}
One can see from here that arrival in groups is less beneficial for the population: the probability of the patch being occupied is maximized when $\nu$ goes to zero.

A more important extension is to account for the Allee effect, by which population biologists call a group of effects leading to a reduction in the per-capita growth rate at small population sizes \cite{Allee}. When the Allee effect is present, a non-zero critical population size for establishment appears already in deterministic theory. If the initial population size is smaller than the critical size, the population goes extinct quickly, whereas if the initial population size is greater than the critical size, a long-lived population appears. A simple way to account for the Allee effect is to modify the reproduction rate $\lambda_n$ in Eq. (\ref{verhulst}), so that
\begin{equation}\label{alleebirth}
\lambda_n= \frac{B\,n}{1+n_0/n},
\end{equation}
where the magnitude of the Allee effect is governed by the parameter $n_0$.   At low immigration rates, and when the population can be long-lived without immigration, the Allee effect changes the quantities $\mu$ and $R$ entering Eq.
(\ref{steadysimple}). The extinction rate of the population without immigration is $\mu=1/\tau_A$, where $\tau_A$
is the mean time to extinction with account of the Allee effect. For the Verhulst model with the modified reproduction rate (\ref{alleebirth}) an accurate approximation for $\tau_A$ can be calculated from the general relations of Ref. \cite{AM}. The result is
\begin{equation}\label{MTE1}
    \tau_A \simeq  \frac{4 \pi \sqrt{B q_0}\,(B-1+Bq_0+D)}{D (B-1-Bq_0+D)^2}\,e^{N\Delta S},
\end{equation}
 where
 \begin{eqnarray}
  \Delta S&=& \frac{D}{B} -\frac{2}{B} \,\mbox{arctanh} \left[\frac{D}{B(1-q_0)+1}\right]\nonumber \\
  &-& 2 q_0\,\mbox{arctanh} \left[\frac{D}{B(1+q_0)-1}\right],
  \label{action}
\end{eqnarray}
$q_0=n_0/N$, and $D=[(B-1-q_0 B)^2-4 B q_0]^{1/2}$.

To evaluate $R$ we should solve the recursion equation (\ref{recursive}), where we can  neglect the density dependence in the mortality rate $\mu_n$ but must keep it in the reproduction rate $\lambda_n$ from Eq.~(\ref{alleebirth}) (the latter procedure is legitimate at $n_0\ll N$). The solution is
\begin{equation}\label{regularized}
p_n=\frac{\left(1-\frac{1}{B}\right)^{n_0+1}
   (n+n_0)!\,
   _2 \tilde{F_1}\left(1,n+ n_0+1;n+1;\frac{1}{B}\right)}{B^n\,n_0!}\,,
\end{equation}
where $_2 \tilde{F}_1$ is the regularized generalized hypergeometric function \cite{mathematica}. For $n=1$, a single immigrant, we obtain a simple formula $p_1=1-\left(1-1/B\right)^{n_0+1}$ (which, in its turn, coincides with our previous result $p_1=1/B$ for $n_0=0$: when the Allee effect is absent).  Therefore, the establishment probability for the single immigrant is $1-p_1=(1-1/B)^{n_0+1}$, and the rate of establishment is $R_A=r(1-1/B)^{n_0+1}$. Finally,
we use Eq.~(\ref{steadysimple}) to predict the stationary probability of observing the population extinct:
$$
P_0=\frac{1}{1+(1-1/B)^{n_0+1}\,r \tau_A}\,,
$$
with $\tau_A$ from Eq.~(\ref{MTE1}). As expected, a strong Allee effect, $n_0\gg 1$, greatly decreases the establishment probability for the single immigrant and increases $P_0$.

\section{Summary}
We determined the stationary PDF of the population size in a variant of the Verhulst model: a simple stochastic population model with independent births, density-dependent deaths and steady immigration. We solved this problem exactly, and found the minimal immigration rate $r$ needed for bringing the probability of observing the population extinct below some fixed level, for example $0.5$. In the limit of low immigration this probability, see Eq. (\ref{finalP0}), is determined by the ratio
of the rate of successful immigration, $R=r(1-1/B)$, and the extinction rate in the same system but without immigration. We have also established, in this limit, a simple relation between the PDF
of the population size in the steady state with immigration and the long-lived quasi-stationary distribution of the population without immigration.  The low-immigration asymptotics enabled us to establish connection between the detailed
stochastic population model (\ref{master}) and the simple patch occupancy model (\ref{set}).  Finally, we have exploited this connection for predicting the relaxation time (\ref{reltime}) of the population size distribution toward its stationary state, and for extending our main results to arrivals in groups and to populations exhibiting the Allee effect.
\vspace{0.5cm}

\section*{Acknowledgements}
We gratefully acknowledge fruitful discussions with Michael Assaf and Pavel V. Sasorov. This work was supported by the Israel Science Foundation (Grant No. 408/08 to B.M.), by the US-Israel Binational Science Foundation (Grant No. 2008075 to B.M.), by the Academy of Finland (Grant No. 250444 to O.O.), and by the European Research Council (ERC Starting Grant No. 205905 to O.O.).

\section*{Appendix}
\renewcommand{\theequation}{A\arabic{equation}}
\setcounter{equation}{0}

Here we simplify Eqs. (\ref{stpdf}) and (\ref{P0}) by employing the strong inequalities  $N\gg 1$ and $r\ll B$ and using asymptotic expansions of the special functions in Eqs.~(\ref{stpdf}) and (\ref{P0}). First, we use an integral representation of the Kummer function \cite{Abramowitz}, and rewrite Eq.~(\ref{P0}) as
\begin{eqnarray}
\label{P0a}
P_0&=&\frac{\Gamma\left(\frac{r}{B}\right)\,\Gamma \left(1+\frac{N-r}{B}\right)}{\Gamma
   \left(1+\frac{N}{B}\right)}\nonumber \\
   &\times &\left[\int_0^1 e^{Nt}\, t^{\frac{r}{B}-1}\, (1-t)^{\frac{N-r}{B}}\,dt\right]^{-1}
\end{eqnarray}
and Eq.~(\ref{stpdf}) as
\begin{eqnarray}
\label{stpdf1}
P_n&=&\frac{N^n \,\Gamma\left(n+\frac{r}{B}\right)\,\Gamma \left(1+\frac{N-r}{B}\right)}{n! \,\Gamma
   \left(n+\frac{N}{B}+1\right)}\nonumber \\
   &\times &\left[\int_0^1 e^{Nt}\, t^{\frac{r}{B}-1}\, (1-t)^{\frac{N-r}{B}}\,dt\right]^{-1}\,.
\end{eqnarray}
At $N\gg 1$ and $r \ll B={\cal O}(1)$ the main contributions to the integral in Eqs. (\ref{P0a}) and (\ref{stpdf1}) come from two well separated regions: the region of
$t \to 0$, where the integrand diverges, and a saddle-point region. The contribution at $t\to 0$ is approximately equal to $B/r$, where we have assumed that $r/B \ll 1/\ln N$. (As will be seen shortly, this strong inequality indeed holds in the interesting region of parameters.)

To evaluate the saddle-point contribution, we rewrite the integrand as
$t^{r/B-1}\, (1-t)^{-r/B} e^{N f(t)}$, where $f(t)=t+B^{-1} \ln(1-t)$.  The saddle point
is at $t_*=1-1/B$; it belongs to the interval $0<t<1$ because $B>1$. The second derivative $f^{\prime\prime}(t_*)=-B$. To extend the integration to the infinite interval $(-\infty,\infty)$, two conditions must hold: $\sqrt{N/B}\gg 1$  and
\begin{equation}\label{cond1}
    1-1/B \gg (NB)^{-1/2}\,.
\end{equation}
Performing the Gaussian integration, and adding the contribution of the $t\to 0$ region, we finally obtain
\begin{eqnarray}\label{integral}
&&\int_0^1 e^{Nt}\, t^{\frac{r}{B}-1}\, (1-t)^{\frac{N-r}{B}}\,dt  \nonumber \\
&\simeq&\frac{B}{r}+\sqrt{\frac{2\pi}{B N}}\,\frac{B}{B-1}\,e^{N\left(1-\frac{1}{B}-\frac{\ln B}{B}\right)} \,.
\end{eqnarray}
As a result, Eq.~(\ref{P0a}) becomes
\begin{equation}\label{P0asymp}
    P_0 \simeq \frac{1}{1+\sqrt{\frac{2\pi}{B N}}\,\frac{r}{B-1}
\,e^{N\left(1-\frac{1}{B}-\frac{\ln B}{B}\right)}}\,.
\end{equation}
Using Eq. ~(\ref{MTE}), we can recast this result as Eq.~(\ref{finalP0}). In its turn, Eq.~(\ref{stpdf1}) for $P_n$ becomes Eq.~(\ref{Pnappr}).


\begin{thebibliography}{99}
\bibitem{rescaled} As other rates in the model, this rate is rescaled by the linear
in $n$ term in the mortality rate.
\end{thebibliography}
\end{document}